\documentclass[11pt,paper,twoside]{JHEP3}

\usepackage{graphicx}                     
\usepackage{amsmath,amssymb,amsfonts}
\usepackage{xspace}                       
\usepackage{float}
\usepackage{dcolumn}
\usepackage{fancyhdr}
\usepackage{citesort}

\newcommand{\beq}{\begin{equation}}
\newcommand{\eeq}{\end{equation}}
\newcommand{\beqa}{\begin{eqnarray}}
\newcommand{\eeqa}{\end{eqnarray}}
\newcommand{\Order}{\mathcal{O}}
\newcommand{\Lagr}{\mathcal{L}}
\newcommand{\BC}{B$\chi$PT\xspace}
\newcommand{\eq}{~=~}

\title{Hyperon decay form factors in chiral perturbation theory\thanks{Work
       supported in part by funds provided from
       the EU to the project ``Study of Strongly Interacting Matter'' under contract no.
       RII3-CT-2004-506078 and by the DFG (SFB/TR 16, ``Subnuclear Structure
       of Matter''), and by the EU Contract No.~MRTN-CT-2006-035482, ``FLAVIAnet''.}}

\author{Andr\'e Lacour,  Bastian Kubis\\
           Helmholtz-Institut f\"ur Strahlen- und Kernphysik (Theorie),
           Universit\"at  Bonn\\
           Nu{\ss}allee 14-16, D-53115 Bonn, Germany \\
           E-mail: \email{lacour@itkp.uni-bonn.de},~\email{kubis@itkp.uni-bonn.de}}
\author{Ulf-G. Mei{\ss}ner\\
         Helmholtz-Institut f\"ur Strahlen- und Kernphysik (Theorie),
         Universit\"at  Bonn\\
         Nu{\ss}allee 14-16, D-53115 Bonn, Germany \\
         and\\
         Institut f\"ur Kernphysik (Theorie), Forschungszentrum J\"ulich,
         D-52425 J\"ulich, Germany\\ E-mail: \email{meissner@itkp.uni-bonn.de}}

\abstract{
We present a complete calculation of the SU(3)-breaking corrections
to the hyperon vector form factors up to $\Order(p^4)$
in covariant baryon chiral perturbation theory.
Partial higher-order contributions are obtained,
and we discuss chiral extrapolations of the vector form factor 
at zero momentum transfer.
In addition we derive low-energy theorems for the
subleading moments in hyperon decays, the weak Dirac radii and the
weak anomalous magnetic moments, up to $\Order(p^4)$.
}
\keywords{Quark Masses and SM Parameters, Weak Decays, QCD, Chiral Lagrangians}
\preprint{HISKP-TH-07/21, FZJ-IKP-TH-2007-23}

\begin{document}

\section{Introduction}
\label{sec:intro}
\setcounter{footnote}{0}

Hyperon semileptonic decays are interesting for various reasons as they
give information on the weak and the strong interactions in the
light quark sector of QCD (for some
recent experimental determinations, see
e.g.\ Refs.~\cite{Affolder:1999pe,AlaviHarati:2001xk,Batley:2006fc,AlaviHarati:2005ev,Wanke:2007rm}).
The transition
matrix elements are parameterized in terms of three vector current and three
axial current form factors. Of these, the so-called vector form factor
at zero momentum transfer, $f_1(0)$, plays a particular role.
Hyperon decay data allow one to extract  $|V_{us} \cdot f_1(0)|^2$, where
$V_{us}$ is one entry of the Cabibbo--Kobayashi--Maskawa matrix.  Deviations
from SU(3) symmetry are expected to be very small because
the Ademollo--Gatto theorem protects $f_1(0)$ from the
leading SU(3) breaking corrections~\cite{Ademollo:1964sr}. Therefore,
precise calculations of the hadronic corrections to $f_1(0)$
appear feasible, resulting eventually in an accurate extraction of $V_{us}$
from hyperon decays
(for a recent analysis of SU(3) breaking effects in hyperon
decays, see Ref.~\cite{Mateu:2005wi} and references therein).

In this paper, we will concentrate on the leading moments of the
vector current form factors in semileptonic hyperon decays.
There are two privileged frameworks for calculating the QCD corrections
to these form factors, namely lattice QCD and
chiral perturbation theory. First exploratory lattice studies are
just becoming available, see
Refs.~\cite{Guadagnoli:2005zs,Sasaki:2006jp,Guadagnoli:2006gj,Lin:2007de}.
The application of chiral perturbation theory  to the semileptonic hyperon
form factors has a longer history, see
Refs.~\cite{Krause:1990xc,Anderson:1993as,Kaiser:2001yc,Villadoro:2006nj},
with partly contradictory or incomplete results:
Ref.~\cite{Krause:1990xc} neglects the mass splitting
in the baryon ground state octet, while Ref.~\cite{Anderson:1993as} 
is erroneous with respect to the signs of certain contributions
and misses some $1/m$ corrections.  Ref.~\cite{Kaiser:2001yc} is purely
confined to the leading-loop contributions.
In Ref.~\cite{Villadoro:2006nj}, in addition the contributions
of dynamical spin-3/2 decuplet intermediate states were considered
in what is known as the small-scale expansion~\cite{Hemmert:1997ye} generalization of 
chiral perturbation theory with baryons, 
which sometimes leads to an improved convergence behavior of the low-energy 
expansion (see e.g.\ Refs.~\cite{Jenkins:1990jv,Jenkins:1991es,Fettes:2000bb}).

Except for the first of these studies, heavy-baryon chiral perturbation theory
was utilized. Recently, a method was established to perform calculations in
baryon chiral perturbation theory (\BC)  in a manifestly covariant
way~\cite{Becher:1999he} (for a recent review discussing also different
formulations of covariant \BC see Ref.~\cite{Bernard:2007zu}).
It is therefore timely to revisit the calculation of the hyperon
vector form factors in that framework.
In what follows, we perform the full one-loop $\Order(p^4)$ calculation
in covariant \BC of hyperon decays, 
which may serve as a check of previous results~\cite{Villadoro:2006nj}
in a different regularization scheme, but in addition provides
partial higher-order corrections useful for a study of the convergence
behavior of the chiral series.
In doing so, we revert to \BC without dynamical decuplet degrees of freedom,
which in the light of surprisingly big effects found in Ref.~\cite{Villadoro:2006nj}
may be considered problematic.  However, we want to concentrate on 
the resummation of higher-order loop effects and therefore defer an even
more involved calculation of the decuplet effects in infrared regularization 
to a later study.

In addition to the already mentioned
vector form factor at zero momentum transfer, we also
calculate further observables such as weak radii and the weak
anomalous magnetic moments. Those observables
have become measurable nowadays \cite{AlaviHarati:2001xk,Batley:2006fc}, and more results
are expected from high-energy colliders in the future.

The paper is organized as follows.
We define the vector form factors and explain their role in semileptonic hyperon decays
in Sect.~\ref{sec:vff}.
In Sect.~\ref{sec:formalism} we present the
chiral Lagrangians necessary for our calculation and discuss the various low-energy constants.
In Sect.~\ref{sec:F10} we present our results for the form factor $f_1$ at vanishing momentum transfer,
confirming findings of Ref.~\cite{Villadoro:2006nj},
and discussing partial higher-order corrections.
As the convergence behavior of SU(3) \BC is known to be problematic,
we investigate various chiral extrapolations in Sect.~\ref{sec:extrapolation}.
In Sects.~\ref{sec:radii} and \ref{sec:moments} we
discuss the weak Dirac radii and the weak anomalous magnetic moments
of semileptonic hyperon decays.
The conclusions are given in Sect.~\ref{sec:conclusions}.

\pagebreak[4]

\section{Vector form factors}\label{sec:vff}

The structure of ground state hyperon decays as
probed by a charged strangeness-changing weak SU(3) vector current
$V^\mu = V_{us} \, \bar{u} \gamma^\mu s $ is parameterized in term of three form factors,
\begin{equation}
\langle B'(p_2) | V^\mu | B(p_1) \rangle = V_{us} \,
\bar{u}(p_2) \left[ \gamma^\mu \, f_1^{BB'}(t) + \frac{i \sigma^{\mu\nu}
q_\nu}{m_1} \, f_2^{BB'}(t) + \frac{q^\mu}{m_1} \,
f_3^{BB'}(t) \right] u(p_1) ~, \label{eq:FFdef}
\end{equation}
with the momentum transfer $q^\mu = p_2^\mu - p_1^\mu$, $t=q^2$,
and $\sigma^{\mu\nu} = i [\gamma^\mu,\gamma^\nu]/2$.
$m_1$ ($m_2$) is the mass of the initial (final) state baryon.
$f_1$ is sometimes referred to as \textit{the} vector form factor,
$f_2$ as the weak magnetism form factor,
and $f_3$ the induced scalar form factor.
The expansion of these form factors at small momentum transfers
defines slope parameters $\lambda_i$ or, in analogy to electromagnetic form factors, \emph{radii},
\beq
f_i(t) \eq f_i(0) \Bigl\{ 1+ \frac{1}{6}\langle r_i^2 \rangle t + \Order(t^2) \Bigr\}
       \eq f_i(0) \Bigl\{ 1+ \lambda_i \frac{t}{m_1^2} + \Order(t^2) \Bigr\} ~,
\label{eq:radii}
\eeq
such that $\lambda_i = m_1^2\langle r_i^2 \rangle/6$.

We consider the
following strangeness-changing $(s \to u)$  decays in the ground-state
baryon octet
\beq
\Lambda  \to p\, \ell^- \bar\nu_\ell \, , ~
\Sigma^0 \to p\, \ell^- \bar\nu_\ell \, , ~
\Sigma^- \to n\, \ell^- \bar\nu_\ell \, , ~
\Xi^-    \to \Lambda\, \ell^- \bar\nu_\ell \, , ~
\Xi^-    \to \Sigma^0 \ell^- \bar\nu_\ell \, , ~
\Xi^0    \to \Sigma^+ \ell^- \bar\nu_\ell \, , ~
\eeq
where the lepton pair $\ell^- \bar\nu_\ell$ can be electronic or muonic.

Vector ($V$) and axial vector ($A$) current contributions do not interfere in the total decay rate
$\Gamma = \Gamma_A+\Gamma_V$, and $\Gamma_V$ is related to the form factors Eq.~\eqref{eq:FFdef}
by~\cite{Garcia:1985xz}
\begin{align}
\Gamma_V \eq G_F^2 \, |V_{us}|^2 \, \frac{\Delta m^5}{60\pi^3} \, \biggl\{&
\biggl[1-\frac{3}{2}\beta + \frac{6}{7}\beta^2 \Bigl(1+\frac{1}{9}m_1^2\langle r_1^2\rangle \Bigr)\biggr]|f_1(0)|^2 \notag\\
& + \frac{6}{7}\beta^2 \Bigl( {\rm Re}\,f_1(0)f_2(0)^* +\frac{2}{3}|f_2(0)|^2 \Bigr)
+ \Order\bigl(\beta^3,m_\ell^2 \bigr) \biggr\}  ~, \label{eq:beta}
\end{align}
where $\beta=\Delta m/m_1 = (m_1-m_2)/m_1$, $G_F$ is the Fermi constant,
and $m_\ell$ denotes the lepton mass, $\ell=e,\,\mu$.
We note that the induced scalar form factor $f_3$ is suppressed by $m_\ell^2$ and can safely
be neglected at least in the electron channel; we will not consider $f_3$ any further in this work.
The expansion in the small quantity $\beta$ in Eq.~\eqref{eq:beta} demonstrates that
the decay width is dominated by $f_1(0)$, and that subleading contributions
are given by the Dirac radius $\langle r_1^2\rangle$ as well as by the weak magnetism
form factor at vanishing momentum transfer, $f_2(0)$.
Both of these subleading moments will hence be discussed in the following.

In the SU(3) limit the vector form factors at zero momentum
transfer $f_1(0)$ are fixed by the conservation of the SU(3)$_V$
charge. The Ademollo--Gatto theorem~\cite{Ademollo:1964sr} asserts
that SU(3) breaking effects only start at second order in the
symmetry breaking term $(m_s-\hat{m})$,
\begin{equation}
\label{sym_br}
 f_1(0) = f_1^{\rm SU(3)}(0) + \Order \left( (m_s-\hat{m})^2 \right) ~,
\end{equation}
with the average small quark mass $\hat{m} = (m_u + m_d)/2$.
$ f_1^{\rm SU(3)}(0) \equiv g_V$ are the vector couplings in the symmetry limit, which read:
\beq
  \label{su3chargs}
  g_V^{\Lambda p}      = - \sqrt{\frac{3}{2}} ~, ~~
  g_V^{\Sigma^0 p}     = - \frac{1}{\sqrt{2}} ~, ~~
  g_V^{\Sigma^- n}     = - 1                  ~, ~~
  g_V^{\Xi^- \Lambda}  =   \sqrt{\frac{3}{2}} ~, ~~
  g_V^{\Xi^- \Sigma^0} =   \frac{1}{\sqrt{2}} ~, ~~
  g_V^{\Xi^0 \Sigma^+} =   1 ~.
\eeq
Since isospin breaking effects are much smaller than SU(3) breaking effects
($m_d-m_u \ll m_s-\hat{m}$), we neglect the former.
Isospin symmetry then relates the transitions $\Sigma^0 \to p$ and $\Sigma^- \to n$,
as well as $\Xi^- \to \Sigma^0$ and $\Xi^0 \to \Sigma^+$ in a trivial manner:
when dividing by the overall vector charge $g_V$, the corresponding form factors
are equal, hence the number of independent processes reduces from six to four.

\section{Chiral Lagrangians}
\label{sec:formalism}

We will employ chiral perturbation theory
($\chi$PT)~\cite{Weinberg:1978kz,Gasser:1983yg,Gasser:1984gg}
as the effective theory in the low-energy region of QCD
(for a recent review see e.g.\ Ref.~\cite{Bernard:2006gx}).

The chiral effective pseudo-Goldstone boson Lagrangian to leading
order is given by
\begin{equation}
  \Lagr^{(2)}_\phi ~=~ \frac{F_\pi^2}{4} \langle u_\mu u^\mu + \chi^+ \rangle ~,
\end{equation}
where $u_\mu = i u^\dagger \nabla_\mu U u^\dagger$, $u^2=U$ collects the
Goldstone boson fields in the usual manner, and
$\chi^+ = u \chi^\dagger u + u^\dagger \chi u^\dagger$,
$\chi = 2B \,{\rm diag}(m_u, u_d, m_s)+\ldots$ incorporates the quark masses.
As the notation suggests, we can identify the Lagrangian's normalization constant
with the pion decay constant for the purpose of this study, $F_\pi=92.4$~MeV.

For the effective meson-baryon Lagrangian
we employ basis and notation of Refs.~\cite{Frink:2006hx,Frink:07}
(see the related work in Ref.~\cite{Oller:2006yh}).
At leading order, it reads
\begin{equation}
 \Lagr^{(1)}_{\phi B} ~=~ \langle \bar{B} \bigl(i\gamma^\mu[D_\mu,B] - m B\bigr) \rangle
+ \frac{D/F}{2} \langle \bar{B} \gamma^\mu \gamma_5 [u_\mu , B]_\pm \rangle ~,
\end{equation}
where $B$ is the matrix of the ground state octet baryon fields, $m$
is the average octet mass and $D$ and $F$ are the axial vector coupling
constants (strictly speaking, the parameters appearing in the Lagrangian
refer to the chiral SU(3) limit). Their
numerical values can be extracted from hyperon decays and obey the
SU(2) constraint for the axial vector coupling $g_A=D+F=1.26$;
we use $D=0.80$, $F=0.46$.
The following terms from the baryon-meson Lagrangian at second order
are needed to generate the baryon mass splittings at leading order,
as well as the coupling of (traceless) vector currents:
\beqa
 \Lagr^{(2)}_{\phi B} &=& b_{D/F} \bigl\langle \bar{B} [\chi^+,B]_\pm \bigr\rangle
+ i \, b_{5/6} \bigl\langle \bar{B} \sigma^{\mu\nu} \bigl[[u_\mu, u_\nu], B\bigr]_\mp \bigr\rangle
\nonumber\\
&& \qquad
+ i \, b_7 \langle \bar{B} u_\mu \rangle \sigma^{\mu\nu} \langle u_\nu B \rangle
+ b_{12/13} \bigl\langle \bar{B} \sigma^{\mu\nu} [F_{\mu\nu}^+,B]_\mp \bigr\rangle \,.
\label{eq:L2}
\eeqa
We use the numerical values
$b_5 = 0.23~\rm{GeV}^{-1}\,$,
$b_6 = 0.62~\rm{GeV}^{-1}\,$,
$b_7 = 0.68~\rm{GeV}^{-1}\,$
obtained from resonance saturation estimates~\cite{Meissner:1997hn,Kubis:2000aa}.
To the order we consider here,
the effects of $b_{D/F}$ can always be re-expressed in terms of the physical baryon masses,
for which we employ
$m_N = 0.939$~GeV,
$m_\Lambda = 1.116$~GeV,
$m_\Sigma = 1.193$~GeV, and
$m_\Xi = 1.318$~GeV.
In addition, we will occasionally refer to an average octet baryon mass $m=1.151$~GeV.
Finally, $b_{12/13}$ can at leading order be determined from the
anomalous magnetic moments of proton and neutron.

Only two terms, entering the Dirac radii of the baryons, are
needed from the third order Lagrangian,
\beq
 \Lagr^{(3)}_{\phi B} \eq
d_{51/52} \, \langle \bar{B} \gamma^\mu \bigl[[D^\nu, F_{\mu\nu}^+],B\bigr]_\mp \rangle ~.
\label{eq:L3}
\eeq
$d_{51/52}$ can be determined from the Dirac (or electric) radii
of the nucleons~\cite{Kubis:2000aa,Kubis:1999xb}.
At fourth order seven couplings proportional to a quark mass
insertion contributing to the anomalous magnetic moments are of
relevance:
\begin{align}
\Lagr^{(4)}_{\phi B}  & \eq
\alpha_{1/2} \, \bigl\langle \bar{B} \sigma^{\mu\nu}  \Bigl( \bigl[ [ F_{\mu\nu}^+,B  ], \chi^+  \bigr]_\mp
         + \bigl[ F_{\mu\nu}^+,  [ B,\chi^+  ]_\mp \bigr] \Bigr) \bigr\rangle  \notag \\
~& ~~ \quad
+ \alpha_{3/4}\, \bigl\langle \bar{B} \sigma^{\mu\nu} \Bigl( \bigl[\{ F_{\mu\nu}^+,B \}, \chi^+  \bigr]_\mp
+  \bigl\{ F_{\mu\nu}^+,  [ B,\chi^+  ]_\mp \bigr\} \Bigr) \bigr\rangle  \notag \\
~& ~~ \quad
+ \alpha_5 \, \langle \bar{B} \sigma^{\mu\nu} B \rangle \langle F_{\mu\nu}^+ \chi^+ \rangle
+ \alpha_{6/7} \, \bigl\langle \bar{B} \sigma^{\mu\nu} [F_{\mu\nu}^+, B]_\mp \bigr\rangle \langle \chi^+ \rangle ~.
\label{eq:L4}
\end{align}
The term proportional to $\alpha_5$ vanishes for off-diagonal currents,
and hence for weak decay matrix elements, while
$\alpha_{6/7}$ account for a quark mass renormalization of the magnetic couplings $b_{12/13}$.
The operators scaling with $\alpha_{1-5}$ incorporate explicit
breaking of SU(3) symmetry in the anomalous magnetic moments and
have to be fitted to the baryon octet's anomalous magnetic moments~\cite{Meissner:1997hn,Kubis:2000aa}.
$\Lagr^{(4)}_{\phi B}$ also contains two additional counterterms contributing
to the magnetic (Pauli) \emph{radii}~\cite{Kubis:2000aa}, which however we will not
consider here.

\section{The Dirac form factor at zero momentum transfer}\label{sec:F10}

We will calculate the loop diagrams in a manifestly covariant form,
using infrared regularization~\cite{Becher:1999he};
for the diagrams that are to be considered, see Fig.~\ref{fig:diags}.

\FIGURE[t]{
\includegraphics[width=\linewidth]{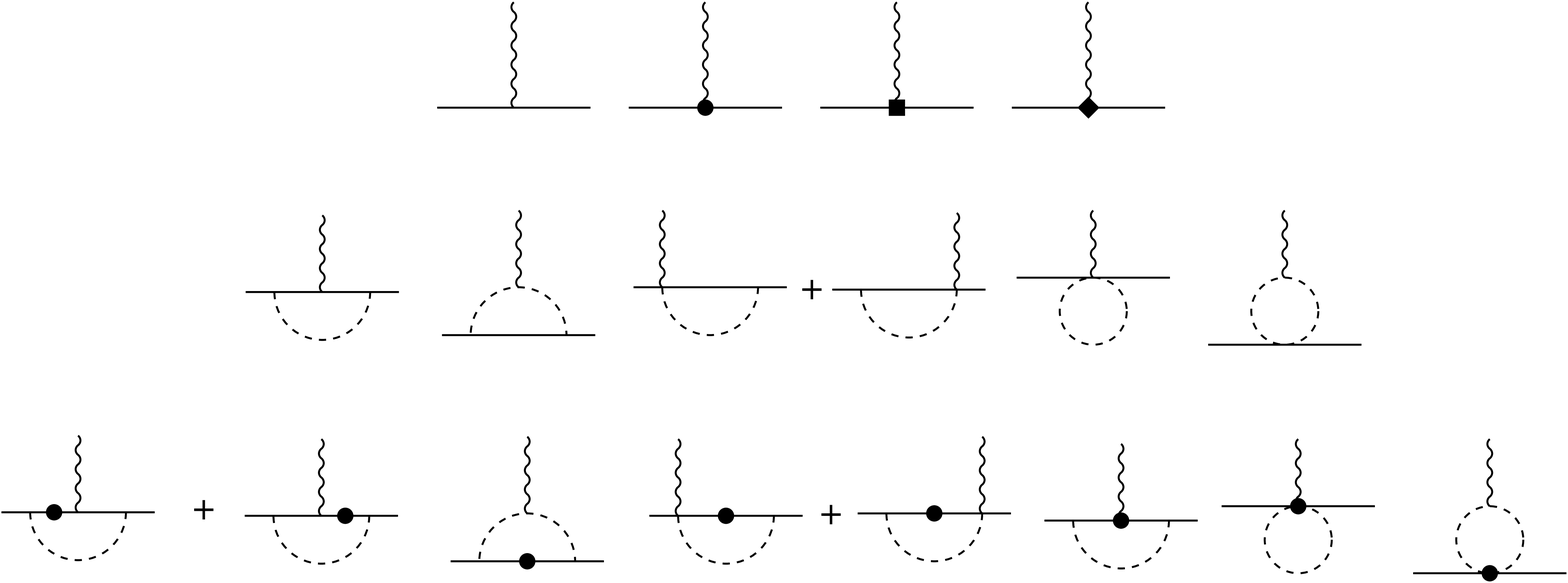}
\caption{
Feynman diagrams contributing to the vector current form factors
up to fourth order. Solid, dashed, and wiggly lines refer to baryons,
Goldstone bosons, and the weak vector source, respectively.
Vertices denoted by a heavy
dot/square/diamond refer to insertions from the
second/third/fourth order chiral Lagrangian, respectively.
Diagrams contributing via wave function renormalization only are not shown.
Note that the masses appearing in the various propagators differ in the
initial and final states and may also be different for the intermediate states.
\label{fig:diags}}
}

In comparison to a heavy-baryon calculation to subleading
one-loop order~\cite{Villadoro:2006nj},
there are far fewer diagrams to be considered, as all the $1/m$ corrections
and baryon mass splittings in the baryon propagators are automatically resummed
to all orders.  On the other hand, the closed forms for the full loop results
are much more involved and cannot be displayed here completely.
A re-expansion of the covariant loop diagrams in strict chiral power counting
however reproduces the heavy-baryon results and leads to simplified expressions
that are useful for comparison with the literature.

The Ademollo--Gatto theorem results in the absence of local contributions
up to fourth chiral order,
therefore our analysis up that order is free from low-energy constants.
We will parameterize the expanded
corrections in analogy to Ref.~\cite{Villadoro:2006nj},
\beq
 f_1(0) \eq g_V \left[ 1 + \delta^{(2)} + \left( \delta^{(3,1/m)} + \delta^{(3,\delta m)} \right)
   + \delta^{(4*)} + \ldots \right]
~.  \label{deff10}
\eeq
$\delta^{(2)}$ is the leading SU(3)-breaking loop correction of order $p^3$.
The corrections of order $p^4$ are divided into two classes,
pure $1/m$ recoil corrections $\delta^{(3,1/m)}$ and
terms proportional to the baryon mass splitting denoted by $\delta^{(3,\delta m)}$.
As an indicator for the size of higher-order terms,
we can also extract \emph{partial} (i.e.\ incomplete)
corrections of order $p^5$ from the covariant amplitudes, which
we denote by $\delta^{(4*)}$ (the asterisk serving as a reminder that there
are additional, e.g.\ two-loop, corrections at that order).

\subsection[Heavy-baryon results up to order $p^4$]{Heavy-baryon results up to order \boldmath{$p^4$}}

Both self-energy like diagrams and tadpoles contribute to $f_1(0)$ at this order,
the former scaling with the axial couplings $D$ and $F$ squared, the latter
coming with completely fixed coefficients.
The results read:
\beq
\delta^{(2)}_{BB'} + \delta^{(3,1/m)}_{BB'}
 \eq 3 \bigl( H^{(1)}_{\pi K} + H^{(1)}_{\eta K} \bigr)
+ \gamma_{BB'}^\pi  \bigl( H^{(1)}_{\pi  K} + H^{(2)}_{\pi K} \bigr)
+ \gamma_{BB'}^\eta \bigl( H^{(1)}_{\eta K} + H^{(2)}_{\eta K}\bigr) ~, \label{eq:delta2+}
\eeq
where the coefficients $\gamma_{BB'}^{\pi/\eta}$ are shown in Table~\ref{tab:coeff},
and the functions $H_{ab}^{(1)}$, $H_{ab}^{(2)}$ are given by
\begin{align}
\label{f1_2o2}
H^{(1)}_{ab} &\eq  \frac{1}{(8 \pi F_\pi)^2}
  \left\{ \frac{M_a^2 M_b^2}{M_b^2-M_a^2} \log \frac{M_b}{M_a} -
  \frac{1}{4} \left(M_a^2+M_b^2\right) \right\} ~, \\
\label{hmcorr}
H^{(2)}_{ab} &\eq \frac{\pi}{6 m(8 \pi F_\pi)^2}
\frac{(M_b - M_a)^2}{M_a + M_b} \left( M_a^2 + 3 M_a M_b + M_b^2 \right) ~.
\end{align}
\TABLE{
\renewcommand{\arraystretch}{1.5}
\begin{tabular}{|c|cc|}
\hline
$B\to B'$ & $\gamma_{BB'}^\pi$ & $\gamma_{BB'}^\eta$ \\
\hline
$\Lambda \to N$  & $9D^2 + 6DF + 9F^2$ & $(D + 3F)^2 $  \\
$ \Sigma \to N$  & $D^2 - 18DF + 9F^2$ & $9(D -  F)^2$ \\
$\Xi \to \Lambda$& $9D^2 - 6DF + 9F^2$ & $(D - 3F)^2 $ \\
$\Xi \to \Sigma$ & $D^2 + 18DF + 9F^2$ & $9(D +  F)^2$ \\
\hline
\end{tabular}
\renewcommand{\arraystretch}{1.0}
\caption{Coefficients for Eq.~\eqref{eq:delta2+}. \label{tab:coeff}}
}

\begin{sloppypar}
The corrections Eqs.~\eqref{f1_2o2}, \eqref{hmcorr}
satisfy the Ademollo--Gatto theorem, and have been given before in the literature
\cite{Krause:1990xc,Kaiser:2001yc,Villadoro:2006nj}.
As already stated in Refs.~\cite{Langacker:1973nf,Gasser:1984ux},
the quadratic symmetry breaking term $(m_s-\hat{m})^2$
comes with coefficients that scale with inverse powers of the
quark masses, therefore allowing for (non-analytic) symmetry-breaking
corrections at lower orders than what local (analytic) terms can provide.
\end{sloppypar}

The baryon mass splitting corrections are somewhat more complicated,
but can still be brought into a rather compact form,
\begin{align}
\delta^{(3,\delta m)}_{\Lambda N} \eq& (D+F)(D+3F) \, H^{\Lambda N}_{K \pi}(m_N) 
- \frac{1}{3}(D^2-9F^2) \, H^{\Lambda N}_{K \eta}(m_N) \notag \\
 &+ \frac{2}{3}D(D+3F) \, H^{\Lambda N}_{\eta  K}(m_\Lambda) + 2D(D-F) \, H^{\Lambda N}_{\pi K}(m_\Sigma) ~, \notag \\
\delta^{(3,\delta m)}_{\Sigma N} \eq& (D^2-F^2) \, H^{\Sigma N}_{K \pi}(m_N) + (D-F)(D-3F) \, H^{\Sigma N}_{K \eta}(m_N) \notag \\
 &- \frac{2}{3}D(D+3F) \, H^{\Sigma N}_{\pi K}(m_\Lambda) - 4(D-F)F \, H^{\Sigma N}_{\pi K}(m_\Sigma) 
  + 2D(D-F) \, H^{\Sigma N}_{\eta K}(m_\Sigma) ~, \notag \\
\delta^{(3,\delta m)}_{\Xi \Lambda} \eq& \frac{2}{3}D(D-3F) \, H^{\Xi \Lambda}_{K \eta}(m_\Lambda) 
 + (D-F)(D-3F) \, H^{\Xi \Lambda}_{\pi K}(m_\Xi) \notag \\
 &- \frac{1}{3}(D^2-9F^2) \, H^{\Xi \Lambda}_{\eta K}(m_\Xi) + 2D(D+F) \, H^{\Xi \Lambda}_{K \pi}(m_\Sigma) ~, \notag \\
\delta^{(3,\delta m)}_{\Xi \Sigma} \eq& - \frac{2}{3}D(D-3F) \, H^{\Xi \Sigma}_{K \pi}(m_\Lambda) 
  + (D^2-F^2) \, H^{\Xi \Sigma}_{\pi K}(m_\Xi) + 4(D+F)F \, H^{\Xi \Sigma}_{K \pi}(m_\Sigma) \notag \\
 &+ (D+F)(D+3F) \, H^{\Xi \Sigma}_{\eta K}(m_\Xi) + 2D(D+F) \, H^{\Xi \Sigma}_{K \eta}(m_\Sigma) ~,
\label{dm_1}
\end{align}
with
\beqa
H^{AB}_{ab}(m) &=& \frac{\pi}{3(8\pi F_\pi)^2} \frac{M_b - M_a}{(M_a + M_b)^2}
\Bigl\{ (m_B-m_A) (M_a^2 + 3 M_a M_b + M_b^2)\nonumber\\
&& \qquad \qquad \qquad \qquad\qquad + 3 (m-m_A) M_b^2 - 3 (m-m_B) M_a^2 \Bigr\} ~,
\label{dm_2}
\eeqa
satisfying the Ademollo--Gatto theorem.
Eq.~\eqref{dm_1} agrees with the results given in Ref.~\cite{Villadoro:2006nj}.

\subsection[Order $p^5$ corrections and numerical results]{Order \boldmath{$p^5$} corrections and numerical results}\label{sec:Op5}

A useful benefit of the infrared regularization method is
that a certain, well-defined subset of higher-order contributions,
stemming from all possible $1/m$ corrections (including, in our case,
those due to the shift of the baryon masses away from their SU(3) symmetry limit)
in the baryon propagators, are automatically resummed.
In the case of the hyperon decay form factors,
such higher-order corrections are far from being complete,
but comprise a complete set of terms, namely those
quadratic in the axial couplings $D$ and $F$.
As a downside, these higher-order terms are in general not finite,
and even after removing the infinities by hand, display some subleading
renormalization scale dependence.
We try to reflect the resulting inherent uncertainties by varying
the scale between $M_\rho=0.770$~GeV and $m_\Xi=1.318$~GeV
(with a central scale chosen at 1~GeV).
Here we evaluate both the partial next-to-next-to-leading order $\delta^{(4*)}$
and the completely resummed covariant loop results numerically.
The analytic expressions are rather cumbersome
and can be obtained in Ref.~\cite{Lacour:070115}.

\TABLE[t]{
\renewcommand{\arraystretch}{1.5}
\begin{tabular}{|c|cccc|ccc|}
\hline
Channel & $\delta^{(2)}$ & $\delta^{(3,1/m)}$ & $\delta^{(3,\delta m)}$ & $\delta^{(4*)}$ & $Sum^{(3)}$ & $Sum^{(4)}$ & $Cov$  \\
\hline
$\Lambda \rightarrow N$   & $-9.7$ & $+8.1$ & $+4.4$ & $+3.8\pm3.2$ & $+2.8$ & $+6.6\pm3.2$ & $-5.7\pm2.1$ \\
$\Sigma  \rightarrow N$   & $+0.8$ & $-3.3$ & $+6.7$ & $+2.3\pm0.6$ & $+4.1$ & $+6.4\pm0.6$ & $+2.8\pm0.2$ \\
$\Xi \rightarrow \Lambda$ & $-6.3$ & $+4.4$ & $+6.3$ & $+0.9\pm2.2$ & $+4.4$ & $+5.3\pm2.2$ & $-1.1\pm1.7$ \\
$\Xi \rightarrow \Sigma$  & $-9.4$ & $+7.9$ & $+2.5$ & $+3.5\pm2.6$ & $+1.0$ & $+4.4\pm2.6$ & $-5.6\pm1.6$ \\
\hline
\end{tabular}
\renewcommand{\arraystretch}{1.0}
\caption{Numerical analysis of relative chiral corrections to the
hyperon decay vector form factors. Values are given for each
contribution separately and by chiral orders summed up
($Sum^{(3)}=\delta^{(2)} + \delta^{(3,1/m)} +
\delta^{(3,\delta m)}$, $Sum^{(4)}=Sum^{(3)}+\delta^{(4*)}$). The
corrections are given in per cent (\%).\label{f10nums}}
}
In Table~\ref{f10nums} we give the numerical results for each
contribution defined in Eq.~\eqref{deff10} ($\delta^{(2)}$,
$\delta^{(3,1/m)}$, $\delta^{(3,\delta m)}$, $\delta^{(4*)}$)
separately, the results summed up to given chiral orders
$\Order(p^3)$--$\Order(p^5)$ ($\delta^{(2)}$, $Sum^{(3)}$,
$Sum^{(4)}$) and for the complete covariant expressions ($Cov$).
Bands for the variation of the renormalization scale as detailed above are
given for $\delta^{(4*)}$, $Sum^{(4)}$, and $Cov$.
The numerical results for $\delta^{(2)}$, $\delta^{(3,1/m)}$, $\delta^{(3,\delta m)}$,
and consequently $Sum^{(3)}$ agree with those of Ref.~\cite{Villadoro:2006nj},
as well as those for $\delta^{(2)}$ with Ref.~\cite{Kaiser:2001yc}.

Comparing with model approaches for the SU(3)-breaking corrections in $f_1(0)$,
we note that quark models tend to yield small negative corrections
of $-1.3\%$~\cite{Donoghue:1986th} or $-2.4\%$ to $-2.5\%$~\cite{Schlumpf:1994fb} 
for all decay channels, while an analysis based on the $1/N_c$ expansion of 
QCD~\cite{FloresMendieta:2004sk} yields positive corrections throughout, ranging
from $+2\pm 2\%$ for $\Lambda\to p$ to $+7\pm 7\%$ for $\Xi \to \Sigma$. 
While heavy-baryon $\chi$PT up to complete $\Order(p^4)$ also favors positive corrections,
the covariant resummation rather agrees in sign with the quark models in three
out of four channels, see Table~\ref{f10nums}.

We note that the $\Sigma \to N$ form factor is consistently
found to receive \emph{positive} corrections in Table~\ref{f10nums},
while there is a sum rule argument~\cite{Quinn:1968qy,Cabibbo:2003cu}
suggesting the opposite.  
The latter is based on the fact that positive contributions 
are due to states of exotic strangeness/isospin quantum numbers
$S=-2$, $I=3/2$, and the assumption that the sum rule ought to be dominated by resonances.
In sum rule language, the corrections calculated in \BC are due to non-resonant
multi-particle intermediate states; as the effects of resonances are subsumed
in counterterms, which only start to contribute at $\Order(p^5)$ for the quantity
at hand, the argument given in Refs.~\cite{Quinn:1968qy,Cabibbo:2003cu}
leads to the expectation that such counterterm corrections are likely to reduce
the positive correction in the $\Sigma \to N$ form factor.

It appears somewhat irritating that the sums of all terms
up to (partial) $\Order(p^5)$, $Sum^{(4)}$, are not at all close
to the full covariant results, see Table~\ref{f10nums},
and even opposite in sign in three out of four channels:
the chiral expansion seems to converge very slowly towards its own
covariant resummation.
This is a frequent problem in SU(3) \BC, see e.g.\
Refs.~\cite{Borasoy:1996bx,Donoghue:1998bs,Kubis:2005cy}.
We further investigate this issue in the following section
by varying the quark masses.

\subsection{Chiral extrapolations}\label{sec:extrapolation}

In order to investigate the convergence behavior of the chiral corrections
to $f_1(0)$ and to test the consistency of our method,
we wish to extrapolate the quark masses towards the chiral limit
and see if convergence in the spirit of chiral power counting
is retrieved for smaller masses.

\FIGURE[t]{
\\
\includegraphics[width=0.48\linewidth]{ccurves_LN.eps} \hfill
\includegraphics[width=0.48\linewidth]{ccurves_SN.eps}
\\[8mm]
\includegraphics[width=0.48\linewidth]{ccurves_XL.eps} \hfill
\includegraphics[width=0.48\linewidth]{ccurves_XS.eps}
\\
\caption{
Chiral extrapolations of the hyperon
decay form factor for the channels $\Lambda \rightarrow N$, $\Sigma \rightarrow N$,
$\Xi \rightarrow \Lambda$, and $\Xi \rightarrow \Sigma$,
according to Eq.~\eqref{eq:ascale}.
The yellow (light) band is the region of the fifth order contribution where we vary
the scale between the $\rho$-- and the $\Xi$--mass.
The green (dark) band is the same variation for the full covariant amplitude.
\label{fig:ce}}
}

To investigate a meaningful quantity that allows for an intuitive
understanding of its convergence properties,
we factor out the symmetry breaking parameter
$(m_s-\hat{m})^2 \propto (M_K^2-M_\pi^2)^2$.
As an extrapolation now would diverge in the chiral limit
due to the appearance of non-analytic quark mass dependence
$\propto m_q^{-1},\,m_q^{-1/2}$,
and in order to obtain a dimensionless quantity,
we additionally factor out a term proportional
to $1/[(4\pi F_\pi)^2(M_\pi^2+M_K^2)]$,
thus making the quark mass dependence of the leading $\Order(p^3)$
symmetry breaking term a simple constant.
We introduce  a quark--mass scaling parameter $a$ and
reparameterize the meson masses according to $M_\phi \rightarrow \sqrt{a} \,M_\phi$.
In Fig.~\ref{fig:ce} we plot the reparameterized
relative corrections to the vector form factor of all four channels,
\begin{equation}
 \Delta(a) \eq \biggl[ \frac{a^2(M_K^2-M_\pi^2)^2}{(4 \pi F_\pi)^2 (M_\pi^2+M_K^2)} \biggr]^{-1}
\left( \frac{f_1(a,t=0)}{{g_V}}-1 \right) ~,
\label{eq:ascale}
\end{equation}
order by order. We observe improved convergence for smaller quark masses
in all four channels, as expected.
As discussed above, the orders $\Order(p^5)$ and $Cov$ include
a subleading dependence on the renormalization scale,
which we vary once more in the region between the $\rho$-- and $\Xi$--mass,
producing bands for these orders.
It is obvious, though, that there is no good convergence
towards the covariant result up to $\Order(p^5)$ even for the
physical quark masses ($a=1$), let alone at even higher masses.

\FIGURE[t]{
\\
\includegraphics[width=0.48\linewidth]{curves2_LN.eps} \hfill
\includegraphics[width=0.48\linewidth]{curves2_SN.eps}
\\[8mm]
\includegraphics[width=0.48\linewidth]{curves2_XL.eps} \hfill
\includegraphics[width=0.48\linewidth]{curves2_XS.eps}
\\
\caption{Chiral extrapolations of the hyperon
decay form factor for the channels $\Lambda \rightarrow N$, $\Sigma \rightarrow N$,
$\Xi \rightarrow \Lambda$, and $\Xi \rightarrow \Sigma$.
We show the dependence on $\hat m/\hat m_{\rm phys}$ for fixed $m_s$.
The yellow (light) band is the region of the fifth order contribution where we vary
the scale between the $\rho$-- and the $\Xi$--mass.
The green (dark) band is the same variation for the full covariant amplitude.
\label{fig:ce2}}
}

As a second application of varying quark masses, and closer in spirit
to what is needed for lattice simulations, we increase the (average)
light quark mass $\hat m$, keeping $m_s$ fixed.
Obviously the convergence of the chiral series becomes rather more
problematic for larger $\hat m$ (or, equivalently, larger pion mass).
However, in the symmetry limit $\hat m = m_s$, or $\hat m/\hat m_{\rm phys} \approx 25$,
the corrections to $f_1(0)$ vanish quadratically,
as can indeed be seen in all channels and for all chiral orders in Fig.~\ref{fig:ce2}.
The qualitative picture is very similar for all channels except the $\Sigma \to N$ one.
Numerically, the covariant results are surprisingly close to the leading-order calculation.
[Note that the bands for the covariant results for the channels $\Sigma \to N$
and $\Xi \to \Sigma$ stop for $\hat m/\hat m_{\rm phys} \lesssim 0.35$, where
the $\Sigma$ becomes instable against strong decay to $\Lambda \pi$,
and $f_1(0)$ develops an imaginary part.]

\subsection{Further comments on the chiral expansion}\label{sec:further}

Let use briefly return to the results of the chiral expansion for the Dirac
form factors at zero momentum transfer displayed in Table~\ref{f10nums}.
First, we note that the corrections taking either the fourth order
heavy-baryon results  $(Sum^{(3)})$ or the full covariant one $(Cov)$ are in the
few percent range, with sizeable  corrections from $1/m$ and baryon mass
splitting insertions in all cases. The importance of these effects makes the
covariant scheme -- in which such contributions are resummed to all orders --
more reliable than the strict heavy-baryon expansion. This is also consistent
with the  experience made in the calculations of the electromagnetic
nucleon~\cite{Kubis:2000zd} and hyperon~\cite{Kubis:2000aa} form factors. It is therefore these
covariant results (last entry in  Table~\ref{f10nums}) that should be used in
the extraction of $V_{us}$ from hyperon semileptonic decays.

\section{Weak radii}
\label{sec:radii}

For the analysis of ongoing and future high-precision hyperon decay experiments,
it will be essential to include the effects of subleading moments in the hyperon
decay vector form factors, first of all the weak magnetic moments $f_2(0)$,
and the leading $t$-dependence of $f_1$
as given by the Dirac radii, see Eq.~\eqref{eq:radii}.
As a side remark, we note that the complete $t$-dependence of the
one-loop chiral representation of $f_1(0)$ displays very little curvature,
such that the linear approximation of Eq.~\eqref{eq:radii} is in fact rather precise;
compare the discussions in Refs.~\cite{Kubis:2000aa,Kubis:2000zd}.

In this section we analyze the weak Dirac radii of the semileptonic hyperon decays.
According to Eq.~\eqref{eq:L3}, two low-energy constants $d_{51/52}$ enter the chiral
representations of these radii at third order (and no further constants at $\Order(p^4)$),
which also feature in the electromagnetic Dirac radii of the nucleons
$\langle r_1^2 \rangle_{p/n}$~\cite{Kubis:2000aa,Kubis:02}.
The hyperon decay radii are therefore linked to the latter by low-energy theorems
that only contain finite, renormalization-scale-independent loop effects as corrections:
\begin{align}
\langle r_1^2 \rangle_{\Lambda N} &\eq \langle r_1^2 \rangle_p + L_r(\Lambda N)   ~, &
\langle r_1^2 \rangle_{\Sigma N}  &\eq \langle r_1^2 \rangle_p +2\,\langle r_1^2 \rangle_n + L_r(\Sigma N) ~,\notag\\
\langle r_1^2 \rangle_{\Xi\Lambda}&\eq \langle r_1^2 \rangle_p + \langle r_1^2 \rangle_n + L_r(\Xi\Lambda) ~, &
\langle r_1^2 \rangle_{\Xi\Sigma} &\eq \langle r_1^2 \rangle_p - \langle r_1^2 \rangle_n + L_r(\Xi\Sigma)  ~.
\label{eq:LETradii}
\end{align}
The remaining loop contributions $L_r(BB')$ are non-analytic SU(3)-symmetry breaking effects.
A representation according to Eqs.~\eqref{eq:LETradii} holds up to corrections of $\Order(p^5)$,
where analytic SU(3) breaking due to quark-mass-dependent counterterms is allowed.
Furthermore, up to $\Order(p^4)$ $f_1(0)$ as factored out in Eq.~\eqref{eq:radii}
can be set to the symmetry limit.
As the neutron Dirac radius is strongly suppressed compared to the proton one
($\langle r_1^2 \rangle_p \simeq 0.60$~fm$^2$ vs.\
$\langle r_1^2 \rangle_n \simeq 0.01$~fm$^2$, see e.g.\ Ref.~\cite{Belushkin:2006qa}), Eqs.~\eqref{eq:LETradii}
suggest hyperon decay radii very close to the proton radius in the SU(3) symmetry limit.

\TABLE[t]{
\renewcommand{\arraystretch}{1.5}
\begin{tabular}{|c|ccc|rrc|}
\hline
Channel & $L_r^{(3)}$ & $L_r^{(4)}$ & $L_r^{(Cov)}$ & $\Order(p^3)$ & $\Order(p^4)$ & $Cov$ \\
\hline
$\Lambda \rightarrow N$   & $-0.24$ & $+0.10$ & $-0.16 \pm 0.06$ & $ 0.35$ & $0.70$ & $0.44 \pm 0.06$ \\
$\Sigma  \rightarrow N$   & $+0.31$ & $+0.10$ & $-0.10 \pm 0.05$ & $ 0.92$ & $0.72$ & $0.51 \pm 0.05$ \\
$\Xi \rightarrow \Lambda$ & $+0.06$ & $+0.05$ & $-0.16 \pm 0.03$ & $ 0.66$ & $0.65$ & $0.45 \pm 0.03$ \\
$\Xi \rightarrow \Sigma$  & $-0.59$ & $+0.19$ & $-0.12 \pm 0.07$ & $-0.01$ & $0.77$ & $0.46 \pm 0.07$ \\
\hline
\end{tabular}
\renewcommand{\arraystretch}{1.0}
\caption{Numerical analysis of the weak Dirac radii squared $\langle r_1^2 \rangle$ 
(in units of fm$^2$).
The columns $L_r^{(3)}$, $L_r^{(4)}$, $L_r^{(Cov)}$ show the SU(3)-breaking loop corrections
as defined in Eq.~\eqref{eq:LETradii} up to third and fourth order, as well as the
full covariant result (the latter with some residual scale dependence).
The last three columns present the resulting Dirac radii.
\label{tab:f1tnums}}
}
The analytic expressions for the loop corrections of the low-energy theorems
tend to be lengthy, these are collected in Appendix~\ref{app:A}.
In Table~\ref{tab:f1tnums} we give the numerical values for these as well as
for the resulting weak Dirac radii in third and fourth
chiral order.
The loop contributions at third order tend to be very large, in particular
for the channel $\Xi \rightarrow \Sigma$, and corrections at fourth order
are similarly sizeable and tend to be opposite in sign, leading to
large cancellations.
For a better assessment of the convergence behavior and a more reliable
prediction, we again show the covariant loop results that consistently
resum all the higher-order $1/m$ and $\delta m$ corrections.
As discussed for the case of $f_1(0)$ above, such partial higher-order
corrections induce some residual dependence on the renormalization scale,
which again we vary around 1~GeV, between the masses of the $\rho$ and the $\Xi$.
After such resummation, all residual loop effects in the radii are negative
and of similar (moderate) size, leading to slightly smaller values
for the hyperon decay radii compared to the proton Dirac radius.
The corresponding numbers are also displayed in Table~\ref{tab:f1tnums}.

We note that phenomenological model parameterizations~\cite{Garcia:1985xz,Cabibbo:2003cu} 
frequently assume a radius term $\langle r_1^2 \rangle = 12/M_V^2 \approx 0.50$~fm$^2$,
derived from a dipole parameterization with $M_V=0.97$~GeV for all
hyperon decay modes, agreeing rather well with the values shown in Table~\ref{tab:f1tnums}.
This agreement is accidental, however, as the SU(3) breaking mechanisms are
completely different:  the dipole mass $M_V$ is obtained by scaling
the phenomenological dipole mass in the nucleon electromagnetic form factors
by $M_{K^*}/M_\rho$, while in $\chi$PT SU(3) breaking is entirely due to 
Goldstone boson loop effects up to the order considered here.  
When resumming the loop contributions in a covariant way, sign and size 
of the effect agrees with the phenomenological guess.

\section{Weak anomalous magnetic moments}
\label{sec:moments}

Comparably to the weak Dirac radii,
the weak magnetism form factor at vanishing momentum transfer $f_2(0)$
can be related by low-energy theorems to anomalous magnetic moments of the
ground state baryon octet.
For convenience, we will define weak magnetic moments with
a slightly different normalization,
\beq
\kappa \eq \frac{2m_N}{m_1} f_2(0) ~,
\eeq
where $m_1$ is the mass of the decaying baryon, see Eq.~\eqref{eq:FFdef},
such that the magnetic moments $\kappa$ are given in units of nuclear magnetons.

At leading order $\Order(p^2)$, the magnetic moments $\kappa_{BB'}$
are just given in terms of the low-energy constants $b_{12/13}$, see Eq.~\eqref{eq:L2},
so they can be strictly related to the proton's and neutron's anomalous magnetic
moments in analogy to Eq.~\eqref{eq:LETradii},
\begin{align}
\kappa_{\Lambda N} &\eq g_V^{\Lambda N} \kappa_p ~, &
\kappa_{\Sigma N} &\eq g_V^{\Sigma N} \bigl( \kappa_p + 2\,\kappa_n \bigr) ~, \notag\\
\kappa_{\Xi\Lambda} &\eq g_V^{\Xi\Lambda} \bigl( \kappa_p + \kappa_n \bigr) ~, &
\kappa_{\Xi\Sigma} &\eq g_V^{\Xi\Sigma} \bigl( \kappa_p - \kappa_n \bigr) ~.\label{eq:LETmmsimple}
\end{align}
Once more $g_V$ are the vector couplings defined in Eq.~\eqref{su3chargs}.
No loop corrections can occur at this order.

However, the large number of symmetry-breaking (i.e.\ quark mass dependent) counterterms
contributing to the magnetic moments at fourth order, see Eq.~\eqref{eq:L4},
prevents a sensible use of a simple low-energy theorem like Eq.~\eqref{eq:LETmmsimple}.
For a discussion of higher-order corrections to such relations,
we therefore resort to relations making use of the well-measured hyperon magnetic moments,
whose chiral representations are also known in covariant \BC
to the appropriate order~\cite{Kubis:2000aa,Kubis:02}.
The corresponding low-energy theorems read
\begin{align}
\label{eq:LETmm}
 \kappa_{\Lambda N}   &\eq g_V^{\Lambda N} \biggl[ \frac{1}{6} \,
\bigl( 5\,\kappa_p + 2\,\kappa_n - 6\,\kappa_\Lambda + \kappa_{\Sigma^-} - 2\,\kappa_{\Xi^-} \bigr)
  + L_\kappa (\Lambda N) \biggr] ~, \notag \\
 \kappa_{\Sigma N}    &\eq g_V^{\Sigma N} \biggl[ \frac{1}{2} \,
\bigl( \kappa_p + 2 \, \kappa_n - \kappa_{\Sigma^+} - 2 \, \kappa_{\Sigma^-} \bigr)
  + L_\kappa(\Sigma N) \biggr] ~, \notag \\
 \kappa_{\Xi \Lambda} &\eq g_V^{\Xi\Lambda} \biggl[ \frac{1}{6} \,
\bigl( 2 \, \kappa_p + 6 \, \kappa_\Lambda - \kappa_{\Sigma^+} - 2 \, \kappa_{\Xi^0} - 5 \, \kappa_{\Xi^-} \bigr)
  + L_\kappa(\Xi \Lambda) \biggr] ~, \notag \\
 \kappa_{\Xi \Sigma}  &\eq g_V^{\Xi\Sigma} \biggl[ \frac{1}{2} \,
\bigl( 2 \, \kappa_{\Sigma^+} + \kappa_{\Sigma^-} - 2 \, \kappa_{\Xi^0} - \kappa_{\Xi^-} \bigr)
  + L_\kappa(\Xi \Sigma) \biggr] ~.
\end{align}
The remaining loop effects $L_\kappa(BB')$ contain non-analytic symmetry-breaking terms.
They are free of fourth-order low-energy constants, and finite up-to-and-including $\Order(p^5)$,
while our calculation of these corrections is complete up-to-and-including
$\Order(p^4)$ only. The corresponding formulae for the $L_\kappa (BB')$
are collected in  Appendix~\ref{app:B}.

\TABLE[t]{
\renewcommand{\arraystretch}{1.5}
\begin{tabular}{|c|ccc|rrrc|}
\hline
Channel & $L_\kappa^{(3)}$ & $L_\kappa^{(4)}$ & $L_\kappa^{(Cov)}$ & $\Order(p^2)$ & $\Order(p^3)$ & $\Order(p^4)$ & $Cov$ \\
\hline
$\Lambda \rightarrow N$   & $-0.16$ & $+0.16$ & $+0.27 \pm 0.03$ & $~~1.33$ & $~~1.17$ & $~~1.49$ & $~~1.60 \pm 0.03$ \\
$\Sigma  \rightarrow N$   & $+0.16$ & $+0.01$ & $+0.03 \pm 0.01$ & $ -1.59$ & $ -1.43$ & $ -1.58$ & $ -1.56 \pm 0.01$ \\
$\Xi \rightarrow \Lambda$ & $-0.16$ & $+0.07$ & $+0.38 \pm 0.05$ & $ -0.13$ & $ -0.29$ & $ -0.07$ & $~~0.25 \pm 0.05$ \\
$\Xi \rightarrow \Sigma$  & $+0.15$ & $+0.25$ & $-0.02 \pm 0.08$ & $~~2.45$ & $~~2.60$ & $~~2.71$ & $~~2.43 \pm 0.08$ \\
\hline
\end{tabular}
\renewcommand{\arraystretch}{1.0}
\caption{Numerical analysis of the weak anomalous magnetic moments (in nuclear magnetons).
The columns $L_\kappa^{(3)}$, $L_\kappa^{(4)}$, $L_\kappa^{(Cov)}$ show the SU(3)-breaking loop corrections
as defined in Eq.~\eqref{eq:LETmm} up to third and fourth order, as well as the
full covariant result (the latter with some residual scale dependence).
The last four columns present the resulting anomalous magnetic moments,
where the $\Order(p^2)$ result is calculated from Eq.~\eqref{eq:LETmm} with vanishing
loop corrections.\label{f20nums}}
}
In Table~\ref{f20nums} we present the numerical results,
both for the residual loop effects and for the resulting
weak anomalous magnetic moments.
All numbers are given to second (no loop effects), third, and fourth chiral order, as
well as the resummation of $1/m$ effects in the covariant loop representation.
The well-known anomalous magnetic moments of the ground state baryons are taken as
$\kappa_p =  1.793$,
$\kappa_n = -1.913$,
$\kappa_\Lambda = -0.613$,
$\kappa_{\Sigma^+} = 1.458$,
$\kappa_{\Sigma^-} = -0.160$,
$\kappa_{\Xi^0} = -1.250$,
$\kappa_{\Xi^-} = 0.349$~\cite{Yao:2006px},
all given in nuclear magnetons.
Again, we show the effects of the residual higher-order scale dependence
in the covariant loop corrections by
varying the renormalization scale between the $\rho$-- and
the $\Xi$--mass.

Comparing third-, fourth-, and covariant loop order results,
we once more observe partially problematic convergence behavior,
with the covariant results not always very close to their
fourth-order approximations.  The total residual loop effects are
very small in two channels ($\Sigma\to N$, $\Xi\to\Lambda$) and more
significant in the other two ($\Lambda\to N$, $\Xi\to\Sigma$).

\section{Conclusions and outlook}\label{sec:conclusions}

Hyperon semileptonic decays allow for an independent method of extracting $V_{us}$,
provided the hadronic form factors involved are sufficiently well-known from theory.
In this article, we have investigated the vector form factors to complete one-loop
order in covariant \BC, with the following findings:
\begin{enumerate}
\item
The dominant contribution in the vector matrix elements of hyperon decays
is due to the Dirac form factor at zero momentum transfer, $f_1(0)$.
Corrections to the SU(3) symmetry limit of this quantity are of second
order in $m_s-\hat m$ due to the Ademollo--Gatto theorem, and are
dominated in the chiral power counting by non-analytic loop contributions.
We confirm earlier results~\cite{Villadoro:2006nj} when performing
the heavy-baryon expansion of the covariant loop results, and find
problematic convergence behavior  of the heavy-baryon series towards
the covariant representation.
\item
An extrapolation towards smaller quark masses confirms that convergence
of the chiral series is restored, but seems problematic at the physical
quark mass values.  The effects of increasing the average light quark mass $\hat m$
at fixed $m_s$ are also presented.
\item
We have argued drawing upon experience made in earlier calculations of nucleon
and hyperon electromagnetic form factors that nevertheless the corrections
obtained from the covariant one-loop result should be utilized in the analysis
of hyperon semileptonic decays.
\item
We have calculated the leading SU(3) breaking corrections to the
radius terms of $f_1(t)$.  These radii can be related to the electromagnetic
Dirac radii of proton and neutron, and corrections are given in terms
of parameter-free loop corrections.  While the convergence behavior is problematic,
the full covariant corrections point at somewhat smaller radii, compared
to the proton radius.
\item
Furthermore, we have established low-energy theorems that relate the weak anomalous magnetic
moments to the magnetic moments of the ground state baryon octet,
valid including leading analytic SU(3) breaking effects.
The non-analytic loop corrections to these relations are also calculated
to complete covariant one-loop order.
\end{enumerate}

\medskip

We have not addressed the much more drastic convergence problems found for the inclusion
of decuplet effects in the calculation of $f_1(0)$~\cite{Villadoro:2006nj}.
A reassessment of these effects in covariant
\BC~\cite{Bernard:2003xf,Bernard:2005fy,Hacker:2005fh,Wies:2006rv}
is an important extension of the present work and will be addressed in a future work~\cite{futurewonders}.

\section*{Acknowledgements}

We would like to thank Matthias Frink for useful discussions.

\appendix

\section{Loop corrections for the weak Dirac radii}
\def\theequation{\Alph{section}.\arabic{equation}}
\setcounter{equation}{0}
\label{app:A}

In this appendix we show analytic expressions for the residual loop contributions to the weak Dirac radii,
$L_r (BB')$, as defined in Eq.~\eqref{eq:LETradii}, to third and fourth order
in the heavy-baryon expansion.
The third order contributions read
\begin{equation}
  \label{eq:tdepO3}
  L_r^{(3)}(BB') =
    \Bigl(\sigma_{BB'}^{\eta K} + \rho_{BB'}^{\eta K}\Bigr) \, H^{(3)}_{\eta K}
  + \Bigl(\sigma_{BB'}^{\pi K}  + \rho_{BB'}^{\pi K}\Bigr) \, H^{(3)}_{\pi  K}
  + \Bigl(\sigma_{BB'}^{K \pi}  + \rho_{BB'}^{K \pi}\Bigr) \, H^{(3)}_{K  \pi}
  ~,
\end{equation}
with
\begin{equation}
H^{(3)}_{ab} \eq \frac{5}{2 (4 \pi F_\pi)^2}
  \left\{ \frac{M_a^4(3M_b^2-M_a^2)}{(M_b^2-M_a^2)^3} \log \frac{M_b}{M_a} -
  \frac{M_a^2 M_b^2}{(M_b^2-M_a^2)^2} + \frac{5}{12} \right\} ~,
\end{equation}
and the coefficients $\sigma_{BB'}^{ab}$, $\rho_{BB'}^{ab}$ are given in Table~\ref{tab:coeff2}.
We decompose the additional fourth order corrections into
$1/m$ recoil corrections and $\delta m$ baryon mass splitting corrections according to
\begin{equation}
  \label{eq:tdepO4}
  L_r^{(4)}(BB') = L_r^{(3)}(BB') + \delta^{(1/m)}_r(BB') + \delta^{(\delta m)}_r(BB') ~.
\end{equation}
The $1/m$ corrections are
\begin{equation}
  \label{eq:tdep1_m}
  \delta^{(1/m)}_r(BB')
  = \rho_{BB'}^{\eta K} \, H^{(4)}_{\eta K}
  + \rho_{BB'}^{\pi K}  \, H^{(4)}_{\pi  K}
  + \rho_{BB'}^{K \pi}  \, H^{(4)}_{K  \pi} ~,
\end{equation}
where
\begin{equation}
  H^{(4)}_{ab} \eq \frac{\pi}{8(4 \pi F_\pi)^2} ~
  \frac{M_a-M_b}{m} ~
  \frac{24M_a^3 + 61M_a^2M_b + 44M_aM_b^2 + 11M_b^3}{(M_a+M_b)^3} ~.
\end{equation}
The $\delta m$ corrections read
\begin{align}
  \label{eq:tdepdm}
  \delta_r^{(\delta m)}(\Lambda N) &\eq (D+F)(D+3F) \, H'^{\Lambda N}_{K
    \pi}(m_N)
  - \frac{1}{3}(D^2-9F^2) \, H'^{\Lambda N}_{K \eta}(m_N) \notag \\
   &~~ \quad + \frac{2}{3}D(D+3F) \, H'^{\Lambda N}_{\eta  K}(m_\Lambda)
  + 2D(D-F) \, H'^{\Lambda N}_{\pi K}(m_\Sigma) \notag \\
   &~~ \quad + \frac{1}{3}(D+3F)^2 H^{(5)}_{\Lambda N} + (D-F)^2 H^{(5)}_{\Sigma N} ~, \notag 
\end{align}
\begin{align}
  \delta_r^{(\delta m)}(\Sigma N) &\eq (D^2-F^2) \, H'^{\Sigma N}_{K \pi}(m_N)
  + (D-F)(D-3F) \, H'^{\Sigma N}_{K \eta}(m_N) \notag \\
   &~~ \quad - \frac{2}{3}D(D+3F) \, H'^{\Sigma N}_{\pi K}(m_\Lambda)
  - 4(D-F)F \, H'^{\Sigma N}_{\pi K}(m_\Sigma) \notag \\
   &~~ \quad + 2D(D-F) \, H'^{\Sigma N}_{\eta K}(m_\Sigma)
  + \frac{1}{3}(D+3F)^2 H^{(5)}_{\Lambda N} + 5(D-F)^2 H^{(5)}_{\Sigma N} ~, \notag \\
  \delta_r^{(\delta m)}(\Xi \Lambda) &\eq \frac{2}{3}D(D-3F) \, H'^{\Xi
    \Lambda}_{K \eta}(m_\Lambda)
  + (D-F)(D-3F) \, H'^{\Xi \Lambda}_{\pi K}(m_\Xi) \notag \\
   &~~ \quad - \frac{1}{3}(D^2-9F^2) \, H'^{\Xi \Lambda}_{\eta K}(m_\Xi)
  + 2D(D+F) \, H'^{\Xi \Lambda}_{K \pi}(m_\Sigma)  \notag \\
   &~~ \quad + \frac{1}{3}(D+3F)^2 H^{(5)}_{\Lambda N} + 3(D-F)^2 H^{(5)}_{\Sigma N} ~, \notag \\
  \delta_r^{(\delta m)}(\Xi \Sigma) &\eq - \frac{2}{3}D(D-3F) \, H'^{\Xi
    \Sigma}_{K \pi}(m_\Lambda)
 + 2D(D+F) \, H'^{\Xi \Sigma}_{K \eta}(m_\Sigma) \notag \\
   &~~ \quad + 4(D+F)F \, H'^{\Xi \Sigma}_{K \pi}(m_\Sigma)
 + (D+F)(D+3F) \, H'^{\Xi \Sigma}_{\eta K}(m_\Xi) \notag \\
   &~~ \quad + (D^2-F^2) \, H'^{\Xi \Sigma}_{\pi K}(m_\Xi)
 + \frac{1}{3}(D+3F)^2 H^{(5)}_{\Lambda N} - (D-F)^2 H^{(5)}_{\Sigma N} ~,
\end{align}
with
\begin{align}
H'^{AB}_{ab}(m) &= \frac{\pi}{(4 \pi F_\pi)^2}
  \biggl\{
 \frac{M_a^2+3M_aM_b+M_b^2}{(M_a+M_b)^3} (m_A+m_B-2m)
- \frac{(M_a-M_b)(m_A-m_B)}{2(M_a+M_b)^2}
\biggl\}  ~,
\nonumber \\
  H^{(5)}_{AB} &= \frac{5\pi}{4(4 \pi F_\pi)^2} \frac{m_A-m_B}{M_K} ~.
\end{align}

\TABLE[t]{
\renewcommand{\arraystretch}{1.5}
\begin{tabular}{|c|cccccc|}
\hline
$B\to B'$ & $\sigma_{BB'}^{\eta K}$ & $\sigma_{BB'}^{\pi K}$ & $\sigma_{BB'}^{K \pi}$
          & $\rho_{BB'}^{\eta K}$ & $\rho_{BB'}^{\pi K}$ & $\rho_{BB'}^{K \pi}$ \\
\hline
$\Lambda \to N$    &
$\frac{3}{5}$      &
$\frac{1}{5}$      &
$\frac{2}{5}$      &
$\frac{1}{3}(D+3F)^2$        &
$(D-F)^2$                    &
$2(D+F)^2$
\\
$ \Sigma \to N$    &
$\frac{3}{5}$      &
$1$                &
$-\frac{2}{5}$     &
$3(D-F)^2$                   &
$\frac{7}{3}D^2-2DF+5F^2$    &
$-2(D+F)^2 $
\\
$\Xi \to \Lambda$  &
$\frac{3}{5}$      &
$\frac{3}{5}$      &
$0$                &
$\frac{1}{3}(D-3F)^2$        &
$3D^2-2DF+3F^2$              &
$0$
\\
$\Xi \to \Sigma$   &
$\frac{3}{5}$      &
$-\frac{1}{5}$     &
$\frac{4}{5}$      &
$3(D+F)^2$                   &
$-\frac{11}{3}D^2-2DF-F^2$   &
$4(D+F)^2$
\\
\hline
\end{tabular}
\renewcommand{\arraystretch}{1.0}
\caption{Coefficients for Eqs.~\eqref{eq:tdepO3}, \eqref{eq:tdepO4}, \eqref{eq:f20O3}.
\label{tab:coeff2}}
}

\section{Loop corrections for the weak anomalous magnetic  moments}
\def\theequation{\Alph{section}.\arabic{equation}}
\setcounter{equation}{0}
\label{app:B}

Here we give the analytic residual loop contributions to the weak anomalous magnetic moments,
$L_\kappa (BB')$, defined in Eq.~\eqref{eq:LETmm}.
The third order contributions can be expressed in terms of the same coefficients
$\rho_{BB'}^{ab}$ given in Table~\ref{tab:coeff2},
\begin{equation}
  \label{eq:f20O3}
  L_\kappa^{(3)}(BB') =
    \rho_{BB'}^{\eta K} \, H^{(6)}_{\eta K}
  + \rho_{BB'}^{\pi K}  \, H^{(6)}_{\pi  K}
  + \rho_{BB'}^{K \pi}  \, H^{(6)}_{K  \pi} ~,
\end{equation}
with
\begin{equation}
  H^{(6)}_{ab} \eq \frac{m \,\pi}{(4 \pi F_\pi)^2} \biggl\{
\frac{M_b^2+M_aM_b-2M_a^2}{3(M_a+M_b)} +
\frac{M_a^2-M_b^2}{2(M_K+M_\pi)} \biggr\}
~.
\end{equation}
The additional fourth order corrections are split into
terms scaling with second-order low-energy constants,
$1/m$ recoil corrections, and $\delta m$ baryon mass splitting corrections,
\begin{equation}
  \label{eq:f20O4}
  L_\kappa^{(4)}(BB') = L_\kappa^{(3)}(BB')
+ \delta^{(4)}_\kappa(BB') + \delta^{(1/m)}_\kappa(BB') + \delta^{(\delta m)}_\kappa(BB') ~.
\end{equation}
These read as follows:
\begin{align}
  \delta^{(4)}_\kappa(\Lambda N) \eq&
   m(3b_5+b_6) \Bigl[ H^{(7)}_{K\pi} + H^{(7)}_{K\eta}  +\frac{8}{3}H^{(8)}_{K\pi} \Bigr] 
 + m\,b_7 \Bigl[ H^{(7)}_{K\eta} - \frac{4}{3} H^{(8)}_{K\pi} \Bigr] 
\notag \\  &+  \biggl\{
\frac{\kappa_p}{4}\biggl[1-\Bigl(\frac{D}{3}+F\Bigr)^2\biggr] 
+\frac{4(\kappa_p+\kappa_n)}{9}\bigl(D^2-3F^2\bigr)  \biggr\} 
\left[ H^{(8)}_{K\pi} + 3 \, H^{(8)}_{K\eta} \right] ~,
\notag \\
  \delta^{(4)}_\kappa(\Sigma N) \eq&
   3m(b_5-b_6) \Bigl[ H^{(7)}_{K\pi} + H^{(7)}_{K\eta}  + \frac{8}{3}H^{(8)}_{K\pi} \Bigr] 
 + m\,b_7 \Bigl[ H^{(7)}_{K\pi} + 4 H^{(8)}_{K\pi} \Bigr] 
\notag \\  &+
\frac{\kappa_p+2\kappa_n}{4}\bigl[1-(D-F)^2\bigr] 
\left[ H^{(8)}_{K\pi} + 3 \, H^{(8)}_{K\eta} \right] ~, \\
  \delta^{(1/m)}_\kappa(\Lambda N) \eq&
   -\frac{1}{2}(3D^2+2DF+3F^2) \, H^{(7)}_{K\pi} - \frac{1}{9}(49D^2+30DF+57F^2) \, H^{(8)}_{K\pi} \notag \\
   &-\frac{1}{6}(D+3F)^2 \, H^{(7)}_{K\eta} + (D^2-2DF-7F^2) \, H^{(8)}_{K\eta} ~, \notag \\
  \delta^{(1/m)}_\kappa(\Sigma N) \eq &
   -\frac{1}{6}(D^2-18DF+9F^2) \, H^{(7)}_{K\pi} 
   + \Bigl(\frac{D^2}{3}+10DF-5F^2\Bigr)  H^{(8)}_{K\pi}  \notag \\
   & -\frac{3}{2}(D-F)^2 \Bigl[ H^{(7)}_{K\eta} +2 H^{(8)}_{K\eta} \Bigr]  ~,
\end{align}
where
\beq
  H^{(7)}_{ab} = \frac{1}{(4 \pi F_\pi)^2} \left\{ \frac{4 \,
      M_b^4}{M_b^2-M_a^2}
    \log \frac{M_b}{M_a} - M_a^2-M_b^2 \right\} ~,~~
  H^{(8)}_{ab} = - \frac{M_b^2 }{(4 \pi F_\pi)^2} \log \frac{M_b}{M_a} ~.
\eeq
The correction terms $\delta^{(4)}_\kappa$ and $\delta^{(1/m)}_\kappa$
for the channels $\Xi \to \Lambda$ and $\Xi \to \Sigma$ can be retrieved
from the corresponding ones in $\Lambda \to N$ and $\Sigma \to N$, respectively, 
by replacing $F \to -F$, $b_6 \to -b_6$, and exchanging
$(\kappa_p+\kappa_n) \leftrightarrow \kappa_p\,$ throughout.
Finally,
\begin{align}
  \label{eq:f20dm}
\delta^{(\delta m)}_\kappa(\Lambda N) &= -(D-F)(D+3F)\,H''^{\Lambda N}_{K \eta}
  - (D^2-6DF-3F^2) \, H''^{\Lambda N}_{K \pi}
  + 2D(D-F) \, H''' \notag \\
 &-(D+F)(D-3F)\, H^{(9)}_{\Lambda N}
  - 2D\biggl[D-F
  -\Bigl(\frac{7}{9}D-F\Bigr) \log \frac{M_\pi}{M_K} \biggr] H^{(9)}_{\Sigma \Lambda} ~, \notag \\
\delta^{(\delta m)}_\kappa(\Sigma N) &= -(D-F)(D+3F)H''^{\Sigma N}_{K \eta}
  + \Bigl(\frac{5}{3}D^2+6DF-5F^2\Bigr)  H''^{\Sigma N}_{K \pi}
  + \frac{2}{3}D(D+3F)  H''' \notag \\
 &+\Bigl(\frac{D^2}{3}+2DF-F^2\Bigr)\, H^{(9)}_{\Lambda N}
  - \biggl[\frac{D^2}{3}+F^2
  - 2D(D+F) \log \frac{M_\pi}{M_K} \biggr] H^{(9)}_{\Sigma \Lambda} ~, \notag \\
\delta^{(\delta m)}_\kappa(\Xi\Lambda) &= (D+F)(D-3F)\,H''^{\Xi\Lambda}_{K \eta}
  + (D^2+6DF-3F^2) \, H''^{\Xi\Lambda}_{K \pi}
  + 2D(D+F) \, H''' \notag \\
 &+(D-F)(D+3F) H^{(9)}_{\Lambda N}
  -\!\biggl[\frac{3}{2}(D^2+F^2) +DF
  -2D\Bigl(\frac{7}{9}D+F\Bigr) \log \frac{M_\pi}{M_K} \biggr] H^{(9)}_{\Sigma \Lambda} \,, \notag \\
\delta^{(\delta m)}_\kappa(\Xi\Sigma) &= (D+F)(D-3F)\,H''^{\Xi\Sigma}_{K \eta}
  - \Bigl(\frac{5}{3}D^2-6DF-5F^2\Bigr) \, H''^{\Xi\Sigma}_{K \pi}
  + \frac{2}{3}D(D-3F) \, H''' \notag \\
 &-\Bigl(\frac{D^2}{3}-2DF-F^2\Bigr)\, H^{(9)}_{\Lambda N}
  - (D-F)\biggl[ \frac{1}{2}(D-F)
  -2D \log \frac{M_\pi}{M_K} \biggr] H^{(9)}_{\Sigma \Lambda} ~,
\end{align}
with
\begin{align}
 H''^{AB}_{ab} &\eq \frac{m}{(4 \pi F_\pi)^2}
  \frac{m_A-m_B}{M_b^2-M_a^2} \biggl\{
  \frac{1}{4}\bigl(M_a^2+M_b^2\bigr) - \frac{M_b^4}{M_b^2-M_a^2}\log\frac{M_b}{M_a}
\biggr\} ~, \\
 H''' &\eq  \frac{m}{(4 \pi F_\pi)^2} \frac{M_\pi^2+M_K^2}{M_\pi^2-M_K^2}\log\frac{M_\pi}{M_K}(m_\Sigma-m_\Lambda)
~, \quad
 H^{(9)}_{AB} \eq \frac{m}{(4 \pi F_\pi)^2}(m_A-m_B) ~. \notag
\end{align}


\end{document}